\documentclass[iopams,floats, floatfix, twocolumn,superscriptaddress, nofootinbib, showpac]{iopart}

\usepackage{graphicx}
\usepackage{xspace} 
\usepackage[usenames,dvipsnames]{color}
\usepackage{dcolumn}
\usepackage{bm}
\usepackage{iopams}
\usepackage{hyperref}
\usepackage{mathrsfs}

\newcommand{\current}{j}

\newcommand{\CITA}{\address{Canadian Institute for Theoretical Astrophysics,
    University~of~Toronto, Toronto, Ontario M5S 3H8, Canada}}

\newcommand{\NYU}{\address{Physics Department, New York University,
    New York, NY, 10003, USA\vspace*{1cm}}}

\begin{document}

\title{Hyperbolicity of Force-Free Electrodynamics} 

\author{Harald P. Pfeiffer}\CITA

\author{Andrew I. MacFadyen }\NYU



We analyze the equations of relativistic magnetized plasma dynamics in
the limiting case that electromagnetic stress-energy is dominant over
pressure and rest mass energy density.  The naive
formulation of these equations is shown to be not hyperbolic.
Modifying the equations by terms that vanish for all physical
solutions, we obtain a symmetric hyperbolic evolution system, which
should exhibit improved numerical behavior.

\section{Introduction}  

Magnetically dominated regions surrounding neutron stars and black
holes, can possess magnetic fields of $\sim 10^{12}$ Gauss (pulsars)
and $\sim 10^{15}$ Gauss (magnetars) and extremely low mass densities
\cite{GoldreichJulian:1969,BlandfordZnajek:1977}.  The magnetic energy
density, $B^2/8\pi$, in these magnetospheres can exceed the rest mass
energy density, $\rho_m c^2$, by many orders of magnitude.  These are
the astrophysical environments responsible for relativistic flows
observed as pulsars, magnetars, quasars and gamma-ray bursts
(GRBs)\cite{Blandford:2002}.  Even in the center of massive stars,
regions of density of $\sim 10^6$ g cm$^{-3}$ are magnetically
dominated for magnetar-like magnetic fields of $10^{15}$ Gauss.

It is therefore of interest to examine the equations governing the
dynamics of relativistic plasma in the limit of extremely strong
magnetic field.  Here, we study the mathematical properties of the
resulting evolutionary equations, with the underlying motivation of
performing computer simulations of astrophysically interesting
situations.

Computatonal codes solving this system have previously been
implemented using a variety of numerical techniques including e.g.
finite difference \cite{Spitkovsky:2006,Contopoulos:2009,
  Palenzuela:2010}, pseudospectral \cite{Petri:2012,Parfrey:2012} and
Godunov schemes
\cite{Komissarov:2002,Asano:2005,Cho:2005,McKinney:2006,Yu:2011}.

We first derive the most immediate formulation of these equations.  We
show that this ``naive'' formulation is not hyperbolic, so that its
initial value problem is not well-posed.  We then modify the evolution
equations and obtain a symmetric hyperbolic evolution system.

\section{Force-Free Condition \& Ohm's Law}

Force-free electrodynamics is a time-evolutionary system
\cite{Uchida:1997,ThompsonBlaes:1998} and can be formulated in terms
of an Ohm's law \cite{Gruzinov:1999}.

The momentum equation for a magnetized fluid can
be written as
\begin{equation}
\rho_m\frac{D\vec{v}}{Dt} = - \nabla P + \rho \vec{E} + \vec\current\times\vec{B}
\end{equation}
where $\rho_m$ is the mass density, $\vec v$ the velocity, $P$ the
pressure, $\rho$ the charge density, $\vec \current$ the current density
and $\vec E$ and $\vec B$ the electric and magnetic field.  We do not
take the non-relativistic limit $(v \ll c)$ so we retain the electric
field in all equations, including the displacement current in
Maxwell's equations (Eq. \ref {eq:Maxwell-E}).  This differs from
non-relativistic magneto-hydrodynamics where the electric field is
neglected.

In the limit that the energy density of the
electromagnetic field is much larger than the fluid pressure and rest
mass energy density $(E^2 + B^2)/8 \pi \gg P, \rho_m $ (we henceforth
use units in which the speed of light $c \equiv 1$) the pressure
gradient term $\nabla P$ and the inertial term
$\rho_m\frac{D\vec{v}}{Dt}$ are negligibly small compared to the
Lorentz force which therefore must also vanish, defining the
force-free condition:
\begin{equation}
\label{eq:forcefree}
\rho \vec{E} +
\vec\current \times \vec{B} = 0.  
\end{equation}

It is assumed that while the mass density of the plasma is small there
are still sufficient charges present at all times to carry any current
necessary to enforce the force-free condition.  In particular, current
is assumed to flow instantaneously along the magnetic field lines to
short out any component of electric field parallel to the magnetic
field, as follows by taking the dot product of
Eq.~(\ref{eq:forcefree}) with $\vec B$:
\begin{equation}\label{eq:EB=0}
\vec E\cdot\vec B=0.
\end{equation}
 We further assume that $B^2-E^2>0$.  If this is not true,
there exists a frame of reference in which $B=0$ and the field is
entirely electric.  This is a regime where effects involving particle
properties will become important and the force-free equations no longer
describe the dynamics.

An expression for the current density in terms of the electric and
magnetic fields can be derived from the force free condition
(Eq. \ref{eq:forcefree}) and Maxwell's equations,
\begin{eqnarray}
\label{eq:Maxwell-E}
& \partial_t\vec E=\nabla\times\vec B-\vec\current,\\
\label{eq:Maxwell-B}
&\partial_t\vec B=-\nabla\times\vec E, \\ 
\label{eq:divE}
&\nabla\cdot\vec E=\rho, \\
\label{eq:divB=0}
& \nabla\cdot\vec B=0,
\end{eqnarray}
as follows.  From Eq.~(\ref{eq:EB=0}), 
\begin{equation}
\label{eq:edotb}
\partial_t{(\vec E\cdot\vec B)} =  \vec E \cdot \partial_t{\vec B} +
\vec B \cdot \partial_t{\vec E} = 0.
\end{equation}

Substituting Maxwell's equations~(\ref{eq:Maxwell-E}) and~(\ref{eq:Maxwell-B})
in Eq.~(\ref{eq:edotb}) yields

\begin{equation}
\label{eq:curls}
\!-\!\vec E\cdot(\nabla\!\times\!\vec E)
\!+\!\vec B\cdot(\nabla\!\times\!\vec B) - \vec B \cdot \vec \current = 0.
\end{equation}

Taking the cross product of the force-free
condition~(Eq. \ref{eq:forcefree}) with $\vec B$, expanding the triple
cross product and using $\rho = \nabla\cdot\vec E$ we get

\begin{equation}
\label{eq:bjb}
\vec B (\vec \current \cdot \vec B) = -(\nabla\cdot\vec E) \vec E \times \vec
B + \vec \current B^2.
\end{equation}

Multiplying Eq. \ref{eq:curls} by $\vec B$ and substituting
Eq.~\ref{eq:bjb} we finally obtain

\begin{equation}\label{eq:j}
\vec\current=\frac{\vec B}{B^2}\left[\vec B\cdot(\nabla\!\times\!\vec B)
\!-\!\vec E\cdot(\nabla\!\times\!\vec E)\right]
+\frac{\vec E\!\times\!\vec B}{B^2}\nabla\cdot\vec E.
\end{equation}

Since all particle properties of the plasma (pressure, inertia) are
assumed negligibly small, the current depends only on the
electromagnetic fields themselves.  The first term on the right-hand
side of Eq. \ref{eq:j} represents the current along the magnetic
field.  We assume perfect conductivity (resistivity $\eta = 0$) so
this term can be non-zero even though $\vec E \cdot \vec B = 0$.  The
second term is the advective current due to charge density ($\rho =
\nabla \cdot \vec E$) moving at the plasma drift velocity.

The current is a non-linear function of the fields which together with
Maxwell's equations Eqs.~(\ref{eq:Maxwell-E}) and~(\ref{eq:Maxwell-B})
yields a set of time-evolution equations for the electromagnetic
fields $\vec E$ and $\vec B$.  We will refer to this set of equations
as the ``naive system,'' and analyze it in the next section.
Physically realistic fields, of course, must always satisfy the
constraints $\vec E\cdot\vec B=0$ and $\nabla\cdot\vec B=0$
(Eqs.~\ref{eq:EB=0} and~\ref{eq:divB=0}).  The $\nabla\cdot\vec E =
\rho$ equation of Maxwell's equations (Eq. \ref{eq:Maxwell-E}) is not
a constraint in force-free electrodynamics, but the definition of
charge density $\rho$.  

\section{Hyperbolicity of the Naive Formulation}

Any system of evolution equations must be well-posed, this means, it
must have a unique solution, and, roughly speaking, small
perturbations of the initial conditions must lead to small
perturbations at later times.  This idea is captured by the
mathematical concept of hyperbolicity (e.g.,
\cite{CourantHilbert:1953,Taylor:1996}).

Denoting the set of evolved fields by $u=\big\{\vec E,\,\vec B\big\}$,
the force-free equations (\ref{eq:Maxwell-E}),~(\ref{eq:Maxwell-B})
with~(\ref{eq:j}) have the structure of a first order system of
evolution equations,
\begin{equation}\label{eq:general-evolution-system}
\partial_t u + A^i\partial_i u=0,
\end{equation}
with matrices $A^i$ depending on the fields $u$ but not their
derivatives.   A system of this form is called {\em strongly hyperbolic}, if,
for each choice of unit-vector $\hat{n}$, the characteristic matrix
\begin{equation}\label{eq:CharacteristicMatrix}
\sum_i \hat n_iA^i
\end{equation}
has all real eigenvalues and a complete set of eigenvectors.  Strong
hyperbolicity is a necessary criterion for
well-posedness\cite{Reula:2004}.

Below, we also use the concept of {\em symmetric hyperbolicity} (see,
e.g. \cite{Taylor:1996}): The evolution system
(\ref{eq:general-evolution-system}) is symmetric hyperbolic if there
exists a positive definite matrix $S$ which simultaneously symmetrizes
all derivative matrices, i.e. $SA^i$ is symmetric for each $i$.  The
{\em symmetrizer} $S$ can depend on the fields $u$ but not their
derivatives.  Symmetric hyperbolicity ensures well-posedness.
Furthermore, because symmetric matrices have all real eigenvectors and
complete sets of eigenvalues, every symmetric hyperbolic system is
also strongly hyperbolic.\footnote{A third concept is {\em strict
hyperbolicity} (e.g. \cite{Taylor:1996}), which asserts that, for
every $\hat n$, the characteristic matrix $\sum_i \hat n_iA^i$ has all
real and distinct eigenvalues.  This implies immediately that the
eigenvectors form a complete set, so that every strictly hyperbolic
evolution system is also strongly hyperbolic.  Strict hyperbolicity is
not applicable for the force-free equations, because the
characteristic speeds are not distinct in all cases.}

The eigenvector analysis of the characteristic matrix
Eq.~(\ref{eq:CharacteristicMatrix}) is not only important for
establishing well-posedness, but also for posing boundary conditions
in numerical simulations.  One must apply boundary conditions
precisely to those characteristic modes that are {\em entering} the
computational domain, for example by the algorithm presented in
\cite{Bjorhus:1995}.

We now examine the eigenvalue problem
\begin{equation}\label{eq:EigenvalueProblem}
\hat n_iA^i e_{(\hat\alpha)}=v_{(\hat\alpha)}\,e_{(\hat\alpha)},
\end{equation}
for the naive force-free equations.  Here, $e_{(\hat\alpha)}$ denotes
the right eigenvectors, labeled by $\hat\alpha=1, \ldots, 6$, and
$v_{(\hat\alpha)}$ are the eigenvalues, or characteristic speeds.

>From Eqs.~(\ref{eq:Maxwell-E}),~(\ref{eq:Maxwell-B}) and~(\ref{eq:j}) we find
\begin{eqnarray} 
A^i
&=\left(\begin{array}{cc}
     A^i_{EE} & A^i_{EB} \\
     A^i_{BE} & 0
 \end{array}\right),
\end{eqnarray}
where each entry represents a $3 \times 3$ submatrix:
\begin{eqnarray}
\left(A^i_{EE}\right)_{jk}
&=-\varepsilon_{ikl}E_lB_j
  +\varepsilon_{jlm}\frac{E_lB_m}{B^2}\delta_{ik},\\
\left(A^i_{EB}\right)_{jk}
&=\varepsilon_{ijk}+\varepsilon_{ikl}\frac{B_j B_l}{B^2},\\
\left(A^i_{BE}\right)_{jk}
&=-\varepsilon_{ijk}.
\end{eqnarray}

Solving Eq.~(\ref{eq:EigenvalueProblem}) results in the characteristic
speeds 
\begin{eqnarray}
\label{eq:v1}
v_{(1)}=&-1,\\
v_{(2)}=&+1,\\
v_{(3)}=&v-w,\\
v_{(4)}=&v+w,\\
v_{(5)}=&v_{(6)}=0,
\label{eq:v6}
\end{eqnarray}
with
\begin{eqnarray}\label{eq:v}
  v&\equiv \frac{\hat n\cdot(\vec E\times\vec B)}{B^2},\\
  w&\equiv \frac{1}{B^2}\sqrt{(\hat n\cdot\vec B)^2(B^2-E^2)}.
\end{eqnarray}

Here, and below, we have used $\vec E\cdot\vec B=0$ to simplify the
expressions.  $v_{(1,2)}$ represent the fast modes, $v_{(3,4)}$ are
the Alfv{\'e}n modes, and $v_{(5,6)}$ are unphysical.  The modes $5$
and $6$ are present because the evolution system has more variables
than physical degrees of freedom, owing to the conditions
Eqs.~(\ref{eq:EB=0}) and (\ref{eq:divB=0}).

We note that $v_{(3,4)}$ become complex when $B^2-E^2<0$, so that in
this regime hyperbolicity is lost.  This reflects the breakdown of the
force-free approximation when $B^2-E^2<0$.

The eigenvectors can most easily be written using the projection
operator orthogonal to $\hat n$; its action on an arbitrary vector
$\vec a$ is defined by 
\begin{equation}
P\vec a\equiv \vec a-(\hat n\cdot\vec a)\hat n.
\end{equation}
Generically, we find 
\begin{eqnarray}
\label{eq:u1}
e_{(1)}&=\left\{-P\vec E+\hat n\!\times\! \vec B,\;\;
P\vec B+\hat n\!\times\!\vec E\right\}^t,\\
\label{eq:u2}
e_{(2)}&=\left\{-P\vec E-\hat n\!\times\! \vec B,\;\;
P\vec B-\hat n\!\times\!\vec E\right\}^t,\\
e_{(3,4)}&=\left\{\;\vec E_{(3,4)},\;\;\vec B_{(3,4)}\right\}^t,\\
\label{eq:u5}
e_{(5)}&=\left\{\;(\hat n\cdot \vec B)\hat n,\; -P\vec E\right\}^t,\\
\label{eq:u6}
e_{(6)}&=\left\{\;\,0,\; \hat n\right\}^t.
\end{eqnarray}
 with
\begin{eqnarray}
\label{eq:AlfvenE}
\vec E_{(3,4)}=&-P\vec B+v_{(3,4)}\hat n\!\times\!\vec E
+\big(1-v_{(3,4)}^2\big)\vec B,\\
\label{eq:AlfvenB}
\vec B_{(3,4)}=&-P\vec E-v_{(3,4)}\hat n\!\times\!\vec B.
\end{eqnarray}

It is interesting to note that because $v_{(1,2)}=\pm 1$, the fast
modes can be written in a form similar in structure to the Alfv{\'e}n
modes. Writing the fast mode eigenvectors as 
$e_{(1,2)}=\left\{\vec E_{(1,2)},
\vec B_{(1,2)}\right\}^t$ with
\begin{eqnarray}
\vec E_{(1,2)}&=-P\vec E-v_{(1,2)}\hat n\times\vec B
+\big(1-v_{(1,2)}^2\big)\vec E,\\
\vec B_{(1,2)}&=+P\vec B-v_{(1,2)}\hat n\times\vec E.
\end{eqnarray}
they have the same form as the Alfv{\'e}n modes
Eqs.~(\ref{eq:AlfvenE}) and~(\ref{eq:AlfvenB}) with the replacements
$\vec E\to\vec B,\; \vec B\to-\vec E$, the duality transformation
between $\vec E$ and $\vec B$.

The left eigenvectors of $\hat n_iA^i$,  defined by 
\begin{equation}
e^{(\hat\alpha)} n_i A^i
=v_{(\hat\alpha)} e^{(\hat\alpha)},\qquad\hat\alpha=1,\ldots,6,
\end{equation}
are given by
\begin{eqnarray}
e^{(1)}&=\left\{ -\vec E+\hat n\times\vec B,\;
                  P\vec B+\hat n\times\vec E\right\},\\
e^{(2)}&=\left\{ -\vec E-\hat n\times\vec B,\; 
                  P\vec B-\hat n\times\vec E\right\},\\
e^{(3,4)}&=\left\{\vec E^{(3,4)},\;\vec B^{(3,4)}\right\},\\
e^{(5)}&= \left\{\vec B,\; P\vec E\right\},\\
e^{(6)}&=\left\{0, \;\hat n\right\},
\end{eqnarray}
with
\begin{eqnarray}
\vec E^{(3,4)}&=
\frac{(P\vec E)^2\!-\!\hat n\!\cdot\!(\vec E\!\times\!\vec B)v_{(3,4)}}
{\hat n\cdot \vec B}
\,\hat n+P\vec B-v_{(3,4)}\hat n\times \vec E,\\
\vec B^{(3,4)}&=P\vec E+v_{(3,4)}\hat n\times\vec B.
\end{eqnarray}
The left and right eigenvectors are orthogonal to each other:
\begin{equation}
e^{(\hat\alpha)}\cdot e_{(\hat\beta)}=0,\qquad \hat\alpha\neq\hat\beta.
\end{equation}
In order to save space, the expressions given above are not normalized.  

\subsection{Breakdown of Hyperbolicity}

Whenever two or more characteristic speeds are equal, it is not
guaranteed that a full set of eigenvectors exists, so those cases must
be examined in detail.  For the naive force-free equations, in many of
these degenerate cases a complete set of eigenvectors does exist, as
detailed in the appendix.

However, in certain cases the eigenvectors are not complete.  For
example, when $\vec E=0$ and $\vec B\cdot\hat n=0$, there exist four
zero-speed eigenvalues.  The eigenvector equation
(\ref{eq:EigenvalueProblem}) reduces in this case to
\begin{eqnarray}
\label{eq:BLA1}
\hat n\times\vec B_{(\hat\alpha)} 
- \hat b \left[\hat b\left(\hat n\times\vec B_{(\hat\alpha)}\right)\right]=0,\\
\label{eq:BLA2}
\hat n\times\vec E_{(\hat\alpha)}=0,
\end{eqnarray}
where $\vec E_{(\alpha)}$ and $\vec B_{(\alpha)}$ denote the electric
and magnetic components of the desired eigenvector, and where $\hat
b\equiv\vec B/B$.  Equations~(\ref{eq:BLA1}) and (\ref{eq:BLA2}) are
solved by $\vec B_{(\hat\alpha)}=C_1\hat n+C_2 \hat n\times\hat b$,
$\vec E_{(\hat\alpha)}= C_3\hat n$ with arbitrary constants
$C_{1,2,3}$, so that the corresponding eigenspace is {\em
three}-dimensional only, and the system is {\em not} strongly
hyperbolic for these values of the variables.

More generally, no complete set of zero-speed eigenvectors exists whenever
(at least) one Alfv{\'e}n-speed vanishes, i.e. when
$v_{(3)}v_{(4)}=0$, or equivalently,
\begin{equation}\label{eq:BLA3}
|\hat n\cdot\vec B|=|P\vec E|.
\end{equation}
Condition (\ref{eq:BLA3}) is very restrictive.  Consider an arbitrary
point in space with values $\vec E$ and $\vec B$ satisfying
$B^2-E^2>0$.  At this point, if $\hat n$ is chosen parallel to $\vec
B$, then $|\hat n\cdot\vec B|-|P\vec E|>0$, whereas for $\hat n$
perpendicular to $\vec B$ we have $|\hat n\!\cdot\!\vec B|-|P\vec
E|\leq 0$.  Therefore, if $\hat n$ changes continuously between these
two directions, Eq.~(\ref{eq:BLA3}) must be satisfied at least once.
At each point in space, no matter what the values of $\vec E$ and
$\vec B$, there exists at least one direction $\hat n$ such that
$\hat n_iA^i$ has no complete set of eigenvectors. Thus, the naive
force-free equations are not strongly hyperbolic.

Komissarov \cite{Komissarov:2002} examined hyperbolicity of a related,
but not identical formulation of force-free electrodynamics.  We
remark that his system behaves similarly to the naive system considered
here: Whenever Eq.~(\ref{eq:BLA3}) holds, no complete set of
eigenvectors exists.

One might argue that even when Eq.~(\ref{eq:BLA3}) holds, there are
``enough'' eigenvectors to represent any physical solution satisfying
$\nabla\cdot\vec B=0$ and $\vec E\cdot\vec B=0$.  Such an observation,
however, is irrelevant because in any numerical simulation, these
constraints will not be satisfied {\em exactly}, but only to truncation
error, or at best to roundoff error.  If the evolution system is not
well-posed, this small constraint-violation may grow on arbitrarily
small timescales.  We conclude that the naive formulation of
force-free electrodynamics is highly unsatisfactory, at best. 

\section{Constraint Addition --- Augmented Evolution System}

The six-dimensional system of force-free dynamics must satisfy the
constraints $\vec E\cdot\vec B=0$ and $\nabla\cdot\vec B=0$
(Eqs.~\ref{eq:EB=0} and \ref{eq:divB=0}).  Addition of terms to the
evolution equations, which are proportional to these constraints, will
not change the physical solutions of the system.  However, if the new
terms contain derivatives, they will modify the $A^i$-matrices and
influence the hyperbolicity of the system.  Our strategy is to add
multiples of such terms to the naive force-free equations, and choose
the coefficients to achieve hyperbolicity.  We augment the naive
force-free equations~(\ref{eq:Maxwell-E}), (\ref{eq:Maxwell-B})
and~(\ref{eq:j}) as follows:
\begin{eqnarray}
\label{eq:dtE+constraints}
\partial_t\vec E=&\nabla\times\vec B-\vec\current
-\gamma_1\frac{\vec E}{B^2}\!\times\!\nabla(E\cdot B),
\\
\label{eq:dtB+constraints}
\partial_t\vec B=&-\nabla\times\vec E
-\gamma_2\frac{\vec E\times\vec B}{B^2}\nabla\cdot\vec B
-\gamma_3\frac{\vec B}{B^2}\!\times\!\nabla(E\cdot B)
\end{eqnarray}
with constants $\gamma_1,\gamma_2,\gamma_3$, and with $\vec \current$
given by (\ref{eq:j}).  The particular form of the new terms was
chosen to have appropriate dimensions and parity, as well as a form
similar to the terms contained in $\vec\current$.  The augmented system
retains the same structure as
Eq.~(\ref{eq:general-evolution-system}).  The choice
$\gamma_1\!=\!\gamma_2\!=\!\gamma_3\!=\!0$ recovers the naive system.

The eigenvalues of the augmented system are
\begin{eqnarray}
\tilde v_{(\hat\alpha)}&=v_{(\hat\alpha)},\qquad\hat\alpha=1,2,3,4,\\
\tilde v_{(5)}&=\gamma_2 v,\\
\tilde v_{(6)}&=(\gamma_3-\gamma_1) v,
\end{eqnarray}
with $v$ given by Eq.~(\ref{eq:v}).  Here, and below we denote
quantities associated with the augmented system with tildes.  The
characteristic speeds $\tilde v_{(1)},\ldots, \tilde v_{(4)}$ are
unchanged, as expected for the physical modes, but those for the
unphysical modes depend on $\gamma_1, \gamma_2$ and $\gamma_3$.

\subsection{Choice of Free Parameters}

We can gain insight into the choices for $\gamma_1,\gamma_2$ and $\gamma_3$ by
considering the zero-speed eigenspace for $\vec E\!=\!0$, $\vec B\cdot\hat
n\!=\!0$, which was found above to be incomplete without constraint
addition (cf. Eqs.~[\ref{eq:BLA1}] and [\ref{eq:BLA2}]).  
The eigenvalue equations for the augmented system reduce in this case to

\begin{eqnarray}\label{eq:BLA4}
\hat n\times\vec B_{(\hat\alpha)} - \hat b \left[\hat b\left(\hat
  n\times\vec B_{(\hat\alpha)}\right)\right]=0,\\
\label{eq:BLA5}
\hat n\times\left[\vec E_{(\hat\alpha)}-\gamma_3\,\hat b\,\left(\vec
b\cdot\vec E_{(\hat\alpha)}\right)\right]=0.\end{eqnarray}

Equation~(\ref{eq:BLA4}) is unchanged by the constraint addition
(cf. Eq.~(\ref{eq:BLA1})) and is solved by $\vec
B_{(\hat\alpha)}=C_1\hat n+C_2\hat n\times\hat b$.  The
structure of Eq.~(\ref{eq:BLA5}) is most interesting: If and only if
$\gamma_3=1$, the square-bracket represents the projection of $\vec
E_{(\hat\alpha)}$ perpendicular to $\hat b$.  Hence, if and only if
$\gamma_3=1$, Eq.~(\ref{eq:BLA4}) has a two-dimensional solution
space, $\vec E_{(\hat\alpha)}=C_3\hat n+C_4\hat b$.  The demand of a
complete set of eigenvectors thus implies $\gamma_3=1$.

The more general case $\vec E$ parallel to $\hat n$, $\vec B\cdot\hat
n=0$ can be dealt with similarly.  The demand of a complete set of
eigenvectors in this case fixes uniquely $\gamma_2=1$.

To fix the remaining parameter $\gamma_1$, consider the
derivative-matrices
\begin{equation}
\tilde A^i=
\left(\begin{array}{cc}
     \tilde A^i_{EE} & \tilde A^i_{EB} \\
     \tilde A^i_{BE} & \tilde A^i_{BB} 
 \end{array}\right).
\end{equation}
For $\gamma_1=0$ (and $\gamma_2=\gamma_3=1$), the off-diagonal blocks
are symmetric, i.e. $\tilde A^i_{EB}=\left(\tilde A^i_{BE}\right)^t$.
Moreover, for $\gamma_1=0$ (and $\gamma_2=\gamma_3=1$), the
characteristic speeds $\tilde v_{(3)}, \ldots, \tilde v_{(6)}$ are
distributed symmetrically around $v$.  Thus, we will choose
\begin{equation}\label{eq:gamma}
\gamma_1=0, \quad\gamma_2=1,\quad\gamma_3=1.
\end{equation}

\subsection{Symmetric Hyperbolicity}

With the choices $\gamma_1=0$, $\gamma_2=\gamma_3=1$, the augmented
system is not only strongly hyperbolic, but even symmetric hyperbolic.
A symmetrizer for this system is given by the $6 \times 6$ matrix
\begin{equation}\fl
S=\frac{1}{B^2}\left(\begin{array}{cc}
  \left(B^2-E^2\right)\delta_{ij}
+\left(z_7\!-\!2\Delta\right)B_iB_j 
& - \Delta\;E_iB_j+ z_7\,B_iE_j  \\[1em]
-\Delta\;B_iE_j+z_7\,E_iB_j
 & \left(B^2-E^2\right)\delta_{ij} + z_7\,E_iE_j
\end{array}\right),
\end{equation}
with $\Delta=1-E^2/B^2$ and $z_7>1$ arbitrary.  The choice $z_7=2$ is
natural, because then $S$ reduces to the identity matrix for $E=0$.
We have not found a symmetrizer for parameters different from
(\ref{eq:gamma}), and therefore believe that this choice is the only one
that makes the augmented system symmetric hyperbolic.

Symmetric hyperbolicity is a very convenient property; in particular,
real eigenvalues and a complete set of eigenvectors are guaranteed.
The remaining part of this section lists these eigenvectors,
beginning with the generic case.  The right eigenvectors of the
physical modes $e_{(1)}, \ldots, e_{(4)}$ are unchanged:
\begin{equation}
\tilde e_{(\alpha)}=e_{(\alpha)},\qquad\alpha=1,2,3,4.\\
\end{equation}
The eigenvectors associated with $\tilde v_{(5)}=\tilde v_{(6)}=v$ are
\begin{eqnarray}
\label{eq:R5}
\tilde e_{(5)}&=\bigg\{
    -\left(B^2-\frac{E^2B_\perp^2}{B^2}\right)\vec B 
    - (\hat n\cdot\vec E)(\hat n\cdot\vec B)\vec E,\nonumber\\
&\qquad\;\; \left(B^2-\frac{E^2B_\perp^2}{B^2}\right)P\vec E 
  - \frac{E^2(\hat n\cdot\vec E)(\hat n\cdot\vec B)}{B^2}P\vec B\bigg\}^t,\\
\label{eq:R6}
\tilde e_{(6)}&=\left\{Q_1 \vec E+Q_2\vec B,\;
Q_3\hat n+Q_4\vec E+Q_5\vec B\right\}^t,
\end{eqnarray}
where
\begin{eqnarray}
Q_1&=B_\perp^2
      \left(\frac{E^2(\hat n\cdot\vec E)^2(\hat n\cdot\vec B)^2}{B^2}
            +B^2(B^2-E^2)(1-v^2)\right),\\
Q_2&=\frac{E^2B_\perp^2(\hat n\cdot\vec E)(\hat n\cdot\vec B)}{B^2}
      \left(B^2-\frac{E^2 B_\perp^2}{B^2}\right),\\
Q_3&=\frac{(B^2-E^2)B_\perp^2}{\hat n\cdot\vec B}
             \left(B^2-E^2+(\hat n\cdot\vec E)^2\right)          
             \left(B^2-\frac{E^2B_\perp^2}{B^2}\right),\\
Q_4&=-\frac{E^2 B_\perp^2(\hat n\cdot\vec E)(\hat n\cdot\vec B)}{B^2}
          \left(B^2-\frac{E^2B_\perp^2}{B^2}\right),\\
Q_5&=\frac{E^2 B_\perp^2}{B^2}\left(B^2-E^2+(\hat n\cdot\vec E)^2\right)
          \left(B^2-\frac{E^2B_\perp^2}{B^2}\right)
\end{eqnarray}
with $B^2_\perp\equiv (P\vec B)^2=(\hat n\times\vec B)^2$.

The left eigenvectors can be written as
\begin{eqnarray}
\tilde e^{(1)}
&=\bigg\{-(\hat n\cdot\vec E) \hat n + (1 + 2v)\hat n\times\vec B
         - \left(v+\frac{(\hat n\cdot\vec B)^2}{B^2}\right)\vec E,\;\nonumber\\
&\qquad\qquad \left(1 + v + \frac{(\hat n\times\vec E)^2}{B^2}\right)\vec B + 
   \hat n\times\vec E - (\hat n\cdot\vec B)\hat n\bigg\},\\
\tilde e^{(2)} 
&=\bigg\{-(\hat n\cdot\vec E)\hat n + (-1 + 2v)\hat n\times\vec B
    + \left(-\frac{(\hat n\cdot\vec B)^2}{B^2}+ v\right)\vec E,\nonumber\\
&\qquad\qquad
    \left(1-v + \frac{(\hat n\times\vec E)^2}{B^2}\right) \vec B - 
          \hat n\times\vec E - (\hat n\cdot\vec B)\hat n\bigg\}, \\
\tilde e^{(3)}
&=\Big\{ (\hat n\cdot\vec B)(B_\perp^2 - E^2)\hat n 
         - (\hat n\cdot\vec B)^2P\vec B - 
          B^2 w\; \hat n\times\vec E, \nonumber\\
&\qquad\qquad
-(\hat n\cdot\vec B)^2 \vec E + B^2 w\; \hat n\times\vec B\Big\},\\
\tilde e^{(4)}
&=\Big\{ -(\hat n\cdot\vec B)(B_\perp^2 - E^2)\hat n 
    + (\hat n\cdot\vec B)^2 P\vec B -  B^2 w\; \hat n\times\vec E, \;\nonumber\\
&\qquad\qquad   (\hat n\cdot\vec B)^2 \vec E + B^2 w\;\hat n\times\vec B\Big\},\\ 
\tilde e^{(5)}&=\left\{\vec B,\; P\vec E\right\},\\
\tilde e^{(6)}&=\left\{0,\;\hat n\right\}.
\end{eqnarray}

We now turn out attention to the degenerate cases:
\begin{enumerate}
\item For $\vec E=\hat n\times\vec B$, $\tilde v_{(1)}=\tilde
v_{(3)}=-1$.  Given any $\hat q$ perpendicular to $\hat n$, this
two-dimensional eigenspace is spanned by
\begin{eqnarray}
\fl \tilde e_{(1)}=Q\left\{\hat q,\;\;-\hat n\times\hat q\right\}^t,\\
\fl \tilde e_{(3)}=\left\{\hat n\times\hat q,\;\;\hat q\right\}^t,\\
\fl \tilde e^{(1)}=\bigg\{-2(\hat n\cdot\vec B)(\hat q\cdot\vec B)\hat n 
+ (\hat n\cdot\vec B)^2\hat q - (\hat q\cdot\vec B)P\vec B, \nonumber\\
\fl \qquad\quad
\left[(\hat n\times\vec q)\cdot\vec B\right]\vec B - B^2\,\hat n\times\hat q\bigg\},\\
\fl \tilde e^{(3)}
=\bigg\{\!-2(\hat n\!\cdot\!\vec B)
    \left[(\hat n\!\times\!\hat q)\!\cdot\!\vec B\right]\hat n 
  + (\hat n\!\cdot\!\vec B)^2\hat n\times\hat q 
  - \left[(\hat n\times\hat q)\!\cdot\!\vec B\right]\,P\vec B, \nonumber\\
\qquad\quad
  -(\vec q\cdot\vec B)\,\vec B + B^2\,\hat q\bigg\}.
\end{eqnarray}

\item For $\vec E=-\hat n\times\vec B$, $\tilde v_{(2)}=\tilde
v_{(4)}=1$.  Given any $\hat q$ perpendicular to $\hat n$, this
two-dimensional eigenspace is spanned by
\begin{eqnarray}
\fl \tilde e_{(2)}&=\left\{\hat q,\;\;\hat n\times\hat q\right\}^t,\\
\fl \tilde e_{(4)}&=\left\{\hat n\times\hat q,\;\;-\hat q\right\}^t,\\
\fl \tilde e^{(2)}&=\bigg\{-2(\hat n\cdot\vec B)(\hat q\cdot\vec B)\hat n 
+ (\hat n\cdot\vec B)^2\hat q - (\hat q\cdot\vec B)P\vec B, \nonumber\\
\fl &\qquad\quad
-\left[(\hat n\times\vec q)\cdot\vec B\right]\vec B + B^2\,\hat n\times\hat q\bigg\},\\
\fl \tilde e^{(4)}
&=\bigg\{\!-2(\hat n\!\cdot\!\vec B)
    \left[(\hat n\!\times\!\hat q)\!\cdot\!\vec B\right]\hat n 
  + (\hat n\!\cdot\!\vec B)^2\hat n\times\hat q 
  - \left[(\hat n\times\hat q)\!\cdot\!\vec B\right]\,P\vec B, \nonumber\\
\fl &\qquad\quad
  +(\hat q\cdot\vec B)\,\vec B - B^2\,\hat q\bigg\}.
\end{eqnarray}
Cases 1. and 2. occur simultaneously if $\vec E=0$ and $\hat n\times\vec
B=0$.  

\item If $\hat n\cdot\vec B=0$, then $\tilde v_{(3)}=\tilde v_{(4)}
=\tilde v_{(5)}=\tilde v_{(6)}=v$.  One can use
\begin{eqnarray}
\tilde e_{(3)}&=
\left\{(1-v^2)B^2\hat n-v(\hat n\cdot\vec E)\hat n\times\vec B,\;
(\hat n\cdot\vec E)\vec B\right\}^t,\\
\tilde e_{(4)}&=\left\{0,\;\hat n\times\vec B\right\}^t,\\
\tilde e_{(5)}&=\left\{\vec B,\;0\right\}^t,\\
\tilde e_{(6)}&=\left\{0,\;\hat n\right\}^t
\end{eqnarray}
and
\begin{eqnarray}
\tilde e^{(3)}&=\left\{\hat n,\;0\right\},\\
\tilde e^{(4)}&=\left\{0,\;\hat n\times\vec B\right\},\\
\tilde e^{(5)}&=\left\{\vec B,\;0\right\},\\
\tilde e^{(6)}&=\left\{0,\;\hat n\right\}.
\end{eqnarray}
This case cannot occur simultaneously with cases 1. or 2. above,
because $\hat n\cdot\vec B=0$ and $\vec E=\pm \hat n\times\vec
B$ imply that $E^2=B^2$, contradicting the assumption
$B^2-E^2>0$.
\item Finally, $\tilde v_{(5)}$ and $\tilde v_{(6)}$ are always equal.
This case has already been incorporated into the general expressions,
Eqs.~(\ref{eq:R5}) and (\ref{eq:R6}).
\end{enumerate}

\section{Conclusion}

We have performed a hyperbolicity analysis of force-free
electrodynamics in the E-B formulation.  The naive evolution system
for this formulation was found to be not hyperbolic, and therefore it
is not well-posed.  We then modified the naive system by addition of
constraints.  The augmented system
Eqs.~(\ref{eq:dtE+constraints})--~(\ref{eq:dtB+constraints}) was shown
to be symmetric hyperbolic for a certain choice of parameters,
$\gamma_1=0, \gamma_2=\gamma_3=1$.  We expect the augmented system to
exhibit better behavior in numerical studies of force-free
electrodynamics.

We thank Lee Lindblom and Mark Scheel for helpful discussions.  HPP
and AIM are grateful for Fairchild and DuBridge fellowships at
Caltech.  This work was supported in part by NSF grants PHY-0244906 to
Caltech, AST-1009863, and the NSERC of Canada.

\section*{Appendix:  Degenerate cases for the naive formulation}

This appendix summarizes those degenerate cases of the naive E-B
system, for which complete sets of eigenvectors exist.
\begin{enumerate}
\item For $\vec E=\hat n\times\vec B$, $v_{(1)}=v_{(3)}=-1$.  Given
any $\hat q$ perpendicular to $\hat n$, this two-dimensional
eigenspace is spanned by
\begin{eqnarray}\label{eq:e_1special}
e_{(1)}&=\left\{\hat q,\; -\hat n\times\hat q\right\}^t,\\
e_{(3)}&=\left\{\hat n\times\hat q,\; \hat q\right\}^t,\\
e^{(1)}&=\left\{\hat q-\frac{\hat q\cdot\vec B}{\hat n\cdot\vec B}\;\hat n,\;
    -\hat n\times\vec q\right\},\\
e^{(3)}&=\left\{\hat n\times\hat q-\frac{(\hat n\times\hat q)\cdot\vec B}
{\hat n\cdot\vec B}\;\hat n,\; \vec q\right\}.
\label{eq:e^3special}
\end{eqnarray}
Note that $\vec E=\hat n\times\vec B$ implies $(\hat n\cdot\vec
B)^2=B^2-E^2>0$.

\item For $\vec E=-\hat n\times\vec B$, $v_{(2)}=v_{(4)}=+1$.  This
case is analogous to $\vec E=\hat n\times\vec B$.  Explicit orthogonal
eigenvectors are given by
Eqs.~(\ref{eq:e_1special})--(\ref{eq:e^3special}) with opposite signs of
the cross-product terms.

\item In the case $\hat n\cdot\vec B=0$, $v_{(3)}=v_{(4)}=v$.  

If $\vec E$ is not parallel to $\hat n$, then $v\neq 0$.  We can
choose
\begin{eqnarray}
e_{(3)}&=\left\{v\vec B,\; \hat n\times \vec B\right\}^t,\\
e_{(4)}&=\left\{\vec E^{(4)},\; \vec B^{(4)}\right\}^t,\\
e^{(3)}&=\left\{P\vec B,\; 0\right\},\\
e^{(4)}&=\left\{\hat n,\;0\right\}
\end{eqnarray}
where
\begin{eqnarray}
\vec E^{(4)}&=\left(1-v^2\right)B\hat n
  -v\frac{\hat n\cdot \vec E}{B}\hat n\times\vec B,\\
\vec B^{(4)}&=\frac{\hat n\cdot \vec E}{B}\vec B.
\end{eqnarray}
If, however, $\vec E$ is parallel to $\hat n$ then $v=0$, leading to a
four-dimensional eigenspace for the eigenvalue $0$.  No complete set
of eigenvectors exists, as discussed in the main text.
\end{enumerate}

\bibliographystyle{unsrt}
\bibliography{forcefree}

\end{document}